# Restarting engine in an X-shaped radio galaxy

**Lakshmi Saripalli**
*Raman Research Institute*
*C. V. Raman Avenue, Sadashivanagar, Bangalore 560080, India*
*E-mail:* `lsaripal@rri.res.in`

**Ravi Subrahmanyan**
*Raman Research Institute*
*C. V. Raman Avenue, Sadashivanagar, Bangalore 560080, India*
*E-mail:* `rsubrahm@rri.res.in`

**Tanmoy Laskar**
*Raman Research Institute*
*C. V. Raman Avenue, Sadashivanagar, Bangalore 560080, India*

**Anton Koekemoer**
*STScI*
*3700 San Martin Drive, Baltimore, MD 21218*
*E-mail:* `koekemoe@stsci.edu`

J2018-556, an X-shaped FR-I radio galaxy provides a unique opportunity to discuss the formation scenarios for the extended structures seen in this class of radio sources. An understanding of the origin of X-shaped source structures has important implications for radio galaxy models as well as black hole-black hole mergers. Guided by our ATCA radio continuum observations of J2018-556 we discuss evidence in support of 'wings' forming as a result of a redirection of backflows by asymmetric gas distribution associated with the host galaxy. Formation models invoking interaction between backflows with asymmetric environments seem the preferred model for X-shaped radio sources despite apparent discoveries of X-shaped FR-I radio sources.

## 1. Introduction

Radio sources having an overall X-shaped structure with twin pairs of lobes centered on a common AGN and axes at large angle with respect to each other are referred to as X-shaped radio sources. One lobe pair is usually brighter than the other. The weaker lobe pair is referred to as 'wings'. The formation of the wings is not understood.

Two contending scenarios for the formation of the wings are: (1) wings form as a result of the deflection of 'back flows' due to an asymmetric gaseous environment around the host galaxy [1,2]. (2) the AGN has undergone two separate epochs of activity. The wings define the axis of a past activity, the brighter lobes define the axis of current activity, and the model requires the black hole to have suffered a large change in its spin axis between the two activity epochs [e.g. 3]. The primary contender for the large axis change is a black hole-black hole merger. In the second model, the abundance of X-shaped sources might be a measure of the event rates for black-hole coalescence signals for gravitational wave detectors [e.g. 3]. However, the spin axis might also change due to perturbations arising in a galaxy interaction; in this case the X-shaped sources might not be a consequence of black-hole coalescence.

A recent study [2] has presented evidence for X-shaped sources to be hosted by galaxies with relatively high ellipticities and the wings to be preferentially aligned closer to the host galaxy minor axis providing strong support for the back-flow scenario. Observation of highly elliptical X-ray gas surrounding the host in an X-shaped radio galaxy 3C403 lends support to this scenario [4].

However, the back-flow origin for X-shaped radio sources requires that back flows exist! Since back flows are known to exist in powerful radio galaxies, the dearth of FR-I type X-shaped sources is viewed as a strong support for the back-flow model [2]. Here we report the case of two FR-I, X-shaped sources, J2018-556 (Fig. 1) and 3C315 [1]. Although of the 'wrong' kind we discuss evidence that strongly supports the back-flow model.

## 2. Formation scenario for J2018-556

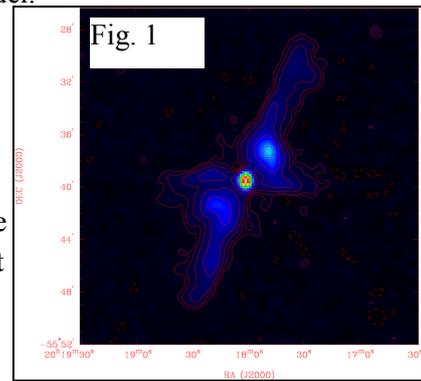

Fig. 1

J2018-556 is the largest known X-shaped source (linear size = 1.53 Mpc, $P_{1.4\,GHz}$ = 1.3×$10^{25}$ W / Hz, $M_R$ = -21.6). J2018-556 is a clear case of an X-shaped source where both lobe pairs have FR-I structure. The properties of the host galaxy and the orientation of host galaxy axes with respect to the two axes of the radio source are same as that found by Capetti et al for X-shaped sources with FR-II structure [2]. Their findings support the model where the wings form as a result of redirection of the back flow in an asymmetric gas distribution encountered. Since the properties of J2018-556 are consistent with the correlations observed for X-shaped FR-II radio galaxies between the host galaxy and the radio structure, it suggests that the wings in J2018-556 and, therefore, its X-shaped structure also have an origin in interactions between back flows and asymmetric gaseous halos.

However, J2018-556 is an FR-I and back flows are not expected to be present in such sources. We propose the following scenario: the giant X-shaped lobes are relics of past activity – during this epoch the source was a powerful FR-II type radio galaxy with back flows that created the wings. The relict giant lobes have relaxed to the current FR-I type structure following cessation of that activity.

That the giant lobes are relicts and were created in a previous FR-II activity phase is also suggested by our discovery that the source has, at its centre, an edge-brightened inner double (Fig. 2): its symmetry, edge-brightened structure and close alignment with the Mpc-size outer double suggest that J2018-556 is being observed at a time when restarted jets are emerging into (relaxed) relict lobes of a past activity phase.

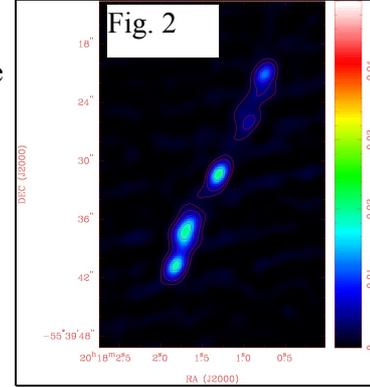

Fig. 2

## 3. Evolutionary history of 3C315

3C315 is a well known X-shaped source and another example where both lobe pairs have FR-I structure. The host galaxy also has high ellipticity and wings close to the minor axis. The remarkable aspect is that just as for J2018-556, in 3C315 also a past activity is inferred independently from the presence of an 8-kpc edge-brightened inner double along the same axis as the main lobe pair.

## 4. Conclusions

J2018-556 is a 1.5 Mpc, FR-I type giant X-shaped radio galaxy. It exhibits the same correlations – of high host ellipticity and alignment of wings with host minor axis – as FR-II type X-shaped sources. Our ATCA continuum observations of J2018-556 resolve the apparent conflict with the back flow scenario with the discovery of a 25-kpc edge-brightened inner double aligned with the outer 1.5 Mpc lobe pair. J2018-556 and 3C315 (also an FR-I X-shape source with an inner double aligned with the outer lobe pair) may both be relics of past FR-II type activity. Formation models invoking interaction between back flows with asymmetric environments are the preferred models for X-shaped radio sources – despite apparent discoveries of X-shaped FR-I radio sources.